\documentclass[final, 4p, times]{elsarticle}
\usepackage{amssymb, amsmath}
\usepackage{textcomp}
\usepackage{subfigure}
\usepackage{braket}
\usepackage{float}
\usepackage{xcolor}
\usepackage{cleveref}
\usepackage{multirow}
\usepackage{icomma}
\usepackage{stackrel}
\usepackage{graphicx}
\usepackage{bm}
\usepackage{appendix}

\def\\cosh{\rm{\cosh}}
\def\\sinh{\rm{\sinh}}

\begin{document}

\title{Nuclear modification factor within a dynamical approach to the complex entropic index}

\author{R. Baptista$^1$, L. Q. Rocha$^1$, J. M. C. Pareja$^1$, T. Bhattacharyya$^2$, A.~Deppman$^1$, E.~Meg\'{\i}as$^{3,*}$,  M. Rybczy\'{n}ski$^2$, G. Wilk$^4$, Z. W\l{}odarczyk$^2$}
 \address{1- Instituto de F\'{i}sica -  Universidade de São Paulo, Rua do Matão 1371, São Paulo 05508-090, Brazil \\ 2- Institue of Physics Jan Kochanowski University, 25-406 Kielce, Poland \\ 3- Departamento de F\'{\i}sica At\'omica, Molecular y Nuclear and  Instituto Carlos I de F\'{\i}sica Te\'orica y Computacional, Universidad de Granada, Avenida de Fuente Nueva s/n,  18071 Granada, Spain \\
 4- National Centre For Nuclear Research, Pasteura 7, Warsaw 02-093, Poland}
\date{*- emegias@ugr.es}

\begin{abstract}
    This work introduces a novel approach to the nuclear deformation factor $R_{\text{AA}}$, grounded in the dynamical effects of the Quark-Gluon Plasma on parton momentum. The approach uses the Blast-Wave method combined with Tsallis Statistics, within the Cooper-Frye freeze-out framework and, by profiting from appropriate simplifications, it gives analytical expressions that describe the observed $R_{\text{AA}}$ for two sets of independent measurements at $\sqrt{s}=2.76$~TeV and $\sqrt{s}=5.02$~TeV. A nonlinear dynamical equation describes the dynamics and leads to log-periodic oscillations. With the analytical solutions for that equation, it is possible to link the dynamical approach with the complex-$q$ formalism, which was proposed to describe the log-oscillations observed in experimental data.
\end{abstract}

\maketitle

\section{Introduction}

One of the most accessible observables in high-energy collisions is the transverse momentum distribution. These distributions are relatively straightforward to measure and offer valuable insight into the underlying collision dynamics. Experimental data consistently show that the overall shape of these distributions is well described by Tsallis statistics, where the probability distribution is proportional to the $q$-exponential function:
\begin{equation}
g(x) \propto e_q(-x) = \left[ 1 + (q-1) x\right]^{\frac{-1}{q-1}}\,,
\end{equation}
with $q > 1$ denoting the entropic index and $x > 1/(1-q)$.


However, a closer look at the observed distributions shows that, in fact, in both $\text{pp}$ and $\text{AA}$ collisions, they are modulated by log-periodic oscillations. Phenomenologically, such oscillations mean that the dynamics of the system being described are richer than those which can be described by the real parameter $q$ (which essentially describes only oscillations and correlations in the system under study). It turns out that it is usually sufficient to enrich the phenomenological description by allowing the entropic index $q$ to take on complex values. A complex entropic index $q$ naturally follows by imposing the requirement of scale invariance in a power law distribution of the form $g(x) = C \cdot x^{-m}$, i.e. $g(\lambda x) = \mu \, g(x)$. Then, it turns out that the generalized $q$-exponential function takes the form $(x \equiv 1 + E / (nT))$~\cite{Wilk2014, Rybczyski2015}
\begin{equation}\label{eq:logoscillation}
g(E) \cong \left[1+\frac{E}{nT} \right]^{-m_0} \left\{w_0 + w_1 \cos \left[ \frac{2\pi}{\ln (1+\alpha)} \ln \left( 1+\frac{E}{nT} \right) \right] \right\}\,, 
\end{equation}
where $m_k = - \ln\mu / \ln\lambda + i 2\pi k / \ln \lambda \; (k = 0, 1, \dots)$, with $\mu \equiv 1-\alpha n$ and  $\lambda \equiv 1+\alpha$. The parameter $\alpha$ plays the role of a scaling factor, and $w_k$ is the weight of order $k$ in a Taylor expansion. $m_0$ approaches $n \equiv \frac{1}{\textrm{Re}(q)-1}$ in the limit $\alpha \to 0$, while $m_k$ has complex values for $k \ge 1$ thus leading to a complex entropic index $q$. In the thermodynamic interpretation, a complex $q$ corresponds to introducing a complex specific heat for the system formed during the collision~\cite{Wilk2014, Rybczyski2015,Wilk2015}, and temperature oscillations associated with sound waves in the hadronic matter~\cite{Wilk2018,Wilk2017-ta}. When a system is subjected to a time-varying temperature, it is continuously driven out of equilibrium and generates entropy. The Fokker-Planck equation can model this behavior, which is what allows for the definition of a complex heat capacity~\cite{Oliveira2019}. The complex heat capacity's real part is the dynamic heat capacity, while its imaginary part is related to the entropy production rate. Solutions to the Fokker-Planck equation provide insights into the internal energy relaxation times, which are related to the eigenvalues of the Fokker-Planck operator.

The goal of our work is to show that the introduction of a complex $q$ can be avoided by a novel approach to calculating the dynamical effects of the expanding medium on the particle momentum distribution. Unifying plasma expansion and hadronisation, this work proposes a simple and effective way to address the dynamics of the quark-gluon plasma. The nuclear modification factor and logarithmic oscillations appear as manifestations of the same dynamical effects. Therefore, the present work contributes to the ongoing efforts to interpret the complex-valued entropic index $q$ by exploring the potential connections with the dynamical aspects of the system formed in high-energy collisions. This novel approach preserves consistency with the nonlinear dynamics resulting from non-local correlations in the medium, while extending the scope of nonextensive methods to dynamical processes in the quark-gluon plasma.

Transverse momentum distributions in nucleus–nucleus collisions differ from those in proton–proton collisions, with a strong dependence on collision centrality, as observed in the nuclear modification factor. In nucleus–nucleus interactions, the system undergoes a centrality-dependent expansion following the formation of the quark–gluon plasma (QGP). During this evolution, numerous partons are produced while the system's temperature decreases. Eventually, the QGP freezes out, leading to the production of a large number of hadrons. This entire process can be broadly divided into three stages: initial particle production, QGP expansion, and freeze-out.

This work analyses the transverse momentum spectra in both nucleus–nucleus and proton–proton collisions, focusing on their differences as quantified by the nuclear modification factor $R_{\text{AA}}$, defined as
\begin{equation}
R_{\text{AA}}(p_T)=\frac{1}{N_{\text{coll}}(\epsilon)}\frac{dN^{AA}/dp_Tdy}{dN^{pp}/dp_Tdy} \,,
\end{equation}
where $dN^{AA}/dp_Tdy$ and $dN^{pp}/dp_Tdy$ are the differential yields in nucleus–nucleus and proton–proton collisions, respectively, and $N_{\text{coll}}(\epsilon)$ is the number of binary nucleon–nucleon collisions for a given centrality $\epsilon$. Our analysis explores the microscopic origin of the complex $q$ parameter and its implications for the dynamical behaviour of parton momentum distributions.

The main objective is to compute the nuclear modification factor by analysing each stage of the reaction individually. To this end, we adopt the following assumptions:
\begin{enumerate}
\item Partons produced in the initial stage follow the same transverse momentum distribution as in proton–proton ($pp$) collisions.
\item The system's expansion is described by the Blast-Wave model, assuming cylindrical symmetry.
\item The in-medium parton dynamics is governed by the non-linear Plastino–Plastino Equation (PPE).
\item The freeze-out surface has cylindrical geometry and, in the region responsible for most observed particles, exhibits axial symmetry.
\end{enumerate}

\section{Theoretical Development}

In the initial stage of the collision, particles are produced via interactions among a subset of partons from the colliding nuclei, with the number of such binary parton-parton collisions denoted by $N_{\text{coll}}(\epsilon)$, which depends on the collision centrality $\epsilon$~\cite{Alice2018, CMS2018}. The transverse momentum distribution of the partons created at this stage is represented by $f(p_T)$.

As the system evolves, collective motion becomes dominant. The produced partons interact strongly, giving rise to a complex, non-linear medium characterized by nonlocal correlations~\cite{Megas2023, Bhattacharyya2023, Bhattacharyya2024}. This expansion stage plays a crucial role in shaping the observed particle distributions.

Eventually, the system hadronizes and undergoes freeze-out, leading to the emission of a large number of hadrons. This process is typically described using the Cooper–Frye formalism, which models particle production through the momentum flux across a three-dimensional hypersurface embedded in four-dimensional spacetime~\cite{Cooper1974}.

\subsection{Cooper–Frye mechanism  in the Blast-Wave picture }

The Cooper–Frye approach assumes that freeze-out occurs when the expanding QGP reaches a three-dimensional surface $\Sigma(x_0,x_1,x_2,x_3)$ in relativistic spacetime. The Lorentz-invariant particle yield is given by
\begin{equation}
E \frac{dN}{d^3p} = \frac{1}{(2\pi)^3}
\int
dN^{\mu} p_{\mu} f^{AA}(p) \,, \label{eq:CooperFrye}
\end{equation}
where $dN^{\mu}$ is the normal vector to the freeze-out hypersurface, and $f^{AA}(p)$ is the particle momentum distribution.

For a longitudinally boost-invariant system, the freeze-out surface can be written as
\begin{equation}
\Sigma_{\mu} = \left( \tau(x,y)\cosh \eta, x, y, \tau(x,y) \sinh \eta \right) \,,
\end{equation}
where $\tau(x,y)$ is the proper time, and $\eta$ is the spacetime rapidity. 
The corresponding normal vector is
\begin{equation}
dN_{\mu} = \left(\cosh \eta,-\frac{\partial \tau}{\partial x},-\frac{\partial \tau}{\partial y},-\sinh \eta \right) \tau(x,y) \, dx dy d \eta \,.
\end{equation}
The four-momentum of a particle is given by
\begin{equation}
p^{\mu}= \left(m_T \cosh \eta', p_x, p_y, m_T \sinh \eta' \right) \,,
\end{equation}
where $\eta'$ is the particle rapidity, and $m_T = \sqrt{p_x^2 + p_y^2 + m^2}$ is the transverse mass. The scalar product between $p^{\mu}$ and $dN_{\mu}$ becomes
\begin{equation}
p^{\mu}\cdot dN_{\mu} =
\left (
m_T \cosh(\eta'-\eta) - p_x \frac{\partial \tau}{\partial x} - p_y \frac{\partial \tau}{\partial y}
\right) \tau \, dx dy d\eta \,.
\end{equation}

The momentum distribution in Eq.~(\ref{eq:CooperFrye}) is generally expressed as a function of the Lorentz-invariant quantity $p^{\mu} u_{\mu}$, where the fluid four-velocity is
\begin{equation}
u_{\mu} = \gamma_v \left( \cosh \eta, -v_x, -v_y, -\sinh \eta \right) \,,
\end{equation}
and $\gamma_v = (1 - v^2)^{-1/2}$ is the Lorentz factor associated with the transverse flow velocity.

When the fluid crosses the freeze-out surface, it hadronizes, and the observed particle spectrum is obtained by integrating over all fluid elements:
\begin{equation}
E \frac{dN}{d^3p} = \frac{1}{(2\pi)^3} \int dx~ dy~ d\eta~ \tau(x,y) \left( m_T \cosh(\eta' - \eta) - p_x \frac{\partial \tau}{\partial x} - p_y \frac{\partial \tau}{\partial y} \right) f(p^{\mu} u_{\mu}) \,.
\end{equation}


In the blast-wave approach, several simplifying assumptions are adopted. First, the fluid expansion is assumed to be axially symmetric~\cite{Grigoryan2017}, allowing the substitution $dx \, dy \rightarrow 2\pi r_T \, dr_T$, where $r_T$ is the radial distance from the symmetry axis. Consequently, the freeze-out proper time becomes a function of the radial coordinate only: $\tau(x,y) = \tau(r_T)$.

Although the proper freeze-out time generally depends on the radial position due to the combined radial and longitudinal expansion of the fluid, in the version of the Blast-Wave model used here, most particles are emitted from a narrow cylindrical shell along the longitudinal direction. This behavior arises from the explosive nature of the system, which imparts momentum to the entire fluid almost instantaneously, resulting in an approximately uniform longitudinal velocity. As a consequence, the freeze-out time can be considered nearly independent of $r_T$, so that $\partial \tau / \partial x = 0$ and $\partial \tau / \partial y = 0$, implying $\tau(r_T) = \tau = \text{const}$. Under this approximation, the QGP freeze-out is considered \emph{quasi-instantaneous}.

Additionally, we assume that the central region of the fluid moves longitudinally with rapidity confined to a narrow interval $\Delta \eta$ around $\eta = 0$. Since most of the detected particles are produced in this region, the integration over the fluid rapidity simplifies the expression for the invariant yield, which becomes:
\begin{equation}
E \frac{dN}{d^3p} = \frac{r_T^2\tau}{8\pi^2}
\Delta \eta \, m_T\cosh \eta^\prime \, f(p^{\mu} u_{\mu}) \,.
\end{equation}

\subsection{Dynamics of partons in the medium}

The partons, while moving through the medium, interact and deposit energy into that medium, and they can reach a quasi-exponential stationary state~\cite{PhysRevLett.84.31}. The fact that the stationary distribution is not the expected Gaussian distribution is indicative of anomalous diffusion, due to the presence of nonlocal correlations leading to the nonextensive statistics. The anomalous diffusion of parton momentum is described by the Plastino–Plastino Equation (PPE), given by~\cite{Plastino1995}
\begin{equation}
\frac{\partial f}{\partial t} - \frac{\partial}{\partial p_i} \left[ A (p_i) f + \frac{\partial}{\partial p_i} \left( D f^{2-q} \right) \right] =0 \,, \label{eq:ppe}
\end{equation}
where $A$ and $D$ are the transport coefficients associated, respectively, with drag and (isotropic) diffusion, and $q$ is a nonextensive parameter.

As the PPE is derived from the Fokker–Planck Equation (FPE) by considering dynamics in fractal spaces, it remains essentially non-relativistic. Relativistic generalisations of the FPE have been proposed~\cite{ChacnAcosta2007} and are widely applied in studies of electromagnetic plasmas. The key difference compared to the PPE is that the time differential operator must be multiplied by the particle energy $E$. However, this simple modification prevents the derivation of analytical solutions. In this work, we adopt the non-relativistic version, acknowledging that predictions for high-momentum partons ($p \gg m$) may underestimate relativistic effects. As will become clear below, such corrections can, in some cases, be effectively absorbed into other parameters of the model. 

It is important to emphasise that the dynamical effects are calculated in the fluid frame, which moves at relativistic velocities when viewed from the laboratory frame. Moreover, the most relevant contributions to the nuclear modification factor stem from the interaction of partons with relatively small momentum in the fluid frame. Indeed, in the Supplementary Material we show that the momentum in the fluid frame is typically about 1/50 of that in the laboratory frame. This explains why many studies have successfully employed the non-relativistic dynamical equation to investigate medium effects in the QGP.

The PPE has been employed to describe the dynamics of partons in the quark–gluon plasma (QGP). This approach is motivated by the fact that the stationary solutions of the PPE exhibit the $q$-exponential form observed in experimental data. While the PPE is valid for a stationary medium, the QGP formed in nucleus–nucleus collisions undergoes rapid expansion as partons propagate through it. Our objective is to investigate how the expanding medium influences the stationary solutions of the PPE.

The PPE calculates dynamical effects in the local fluid rest frame; thus, the relevant momentum for the dynamics is\footnote{In this subsection and in the rest of the manuscript, we omit arrows for 3-vectors. It should be understood that $p \equiv \vec{p}$ and $v \equiv \vec{v}$, unless stated otherwise.}
\begin{equation}
\bar{p}_T^o = L_u[p_T^o] \,,
\end{equation}
where $p_T^o$ is the parton's average transverse momentum at the moment of creation. The main effect of the dynamical processes in the medium is the modification of the parton's momentum, given by
\begin{equation}
\bar{p}_T = \bar{p}_T^o \exp(-A \tau) \,, \label{eq:DragEffect}
\end{equation}
where $\tau$ is the time elapsed between QGP formation and freeze-out. The equation above includes the drag effect. It is also possible to include the diffusion effect; however, it was observed that its effects are negligible when compared with those from drag forces. The parton momentum in the CM frame is obtained by the inverse Lorentz transformation:
\begin{equation}
p_T = L_{-u}[\bar{p}_T] \,.
\end{equation}
In the equations above, $L_{\pm u}$ represents the Lorentz transformation of the CM momentum to ($+$) and from ($-$) the fluid frame.

For convenience, the Lorentz transformations are considered separately. The transformation of the space-like components of the momentum is
\begin{equation}
L_u(\vec{p}) = \gamma_v \left( \vec{p} - \vec{v} E \right) \,,
\end{equation}
and the transformation of the time-like component is
\begin{equation}
L_u(E) = \gamma_v \left( E - \vec{v} \cdot \vec{p} \right) \,,
\end{equation}
where $E = \sqrt{\vec{p}^{,2} + m^2}$ and $u^{\mu} = \gamma_v (1, \vec{v})$ is the local fluid velocity.

The separation between space-like and time-like components is convenient since the dynamical effects are associated solely with the space-like components. To recover the transverse momentum before the drift effects, the following relation is used:~\footnote{By considering the Lorentz transformation of the energy, the expression relating the initial energy $E^o$ with the final energy $E$ of the parton turns out to be $E^o = L_{-u}\left( \sqrt{m^2 + (L_u(E)^2 - m^2 ) \exp(2 A \tau)} \right)$. It is easy to check that this expression agrees with the simpler formula written in Eq.~(\ref{eq:popT}).}
\begin{equation}
\vec{p}_T^o = L_{-u}\left[\exp(A \tau) L_u[\vec{p}_T]\right] \,, \label{eq:popT}
\end{equation}
and the time-like component is calculated to maintain mass invariance,
\begin{equation}
E^o = \sqrt{\vec{p}_T^{o \, 2} + m^2} \,,
\end{equation}
representing the parton energy before drift effects. With $E^o$ and $\vec{p}_T^o$, one forms the full four-momentum of the created parton with mass $m$: $p^{o \, \mu} = (E^o, \vec{p}_T^{o})$.

It is assumed that the momentum distribution at the moment of parton creation has the same functional form for $pp$ and $AA$ collisions, and is described by~\cite{Rocha2022}
\begin{equation}
E\frac{d N}{d^3p} = \frac{gV m_{\text{T}} \cosh{\eta'}}{(2\pi)^3} f(p_T) \,, \label{eq:InvCross-sec}
\end{equation}
where
\begin{equation}
f(p_T) = N \left[ 1 + (q-1)\frac{m_T \cosh \eta' - \mu}{T} \right]^{-\frac{q}{q-1}} \,, \label{eq:multi-distr}
\end{equation}
with $q = 1.16$, $T$ being the temperature, $g$ the degeneracy factor, $N$ the particle multiplicity, and $V$ the system volume. Note that from Eqs.~(\ref{eq:InvCross-sec}) and (\ref{eq:multi-distr}), the distribution satisfies $dN/d^3p \propto N/E$.

In $AA$ collisions, this distribution is deformed due to interactions with the expanding medium. For a given observed momentum $p_T$, the modified distribution $f^{AA}(p_T)$ becomes
\begin{equation}
f^{AA}(p_T) = f(p_T^o) = f\left( L_{-u}\left[ \exp(A \tau), L_u[p_T] \right] \right) \,,
\end{equation}
where Eq.~(\ref{eq:popT}) has been used.

\subsection{Particle multiplicity and nuclear modification factor}

Multiparticle production is inherently a non-extensive process. One of its key consequences is the non-linear relationship between the total collision energy and the produced particle multiplicity. This relationship can be expressed as
\begin{equation}
N \propto E^{1-d} \,,
\label{multsc}
\end{equation}
where $d$ is the fractal dimension of the QGP, expected to be $d = 0.69$~\cite{Deppman2020b}. From this scaling, the ratio of multiplicity to energy follows
\begin{equation}
\frac{N}{E} \propto E^{-d} \,. \label{eq:nonext}
\end{equation}

This behaviour indicates that the system formed in high-energy collisions is non-extensive, since the total energy does not scale linearly with the number of particles. Importantly, the total energy depends on the collision centrality, which is characterised by the fraction of nucleons participating in the overlap region of the colliding nuclei. It is therefore reasonable to assume that the effective energy available for particle production scales as
\begin{equation}
E \propto N_{AA}(1 - \epsilon) \,,
\end{equation}
where $\epsilon$ denotes the eccentricity, expressed in percent, and $N_{AA}$ is the number of participant nucleons. The equation above states that the total energy involved in the collision depends on its centrality.

\begin{figure}[H]
  \centering
  \includegraphics[width=0.3\linewidth]{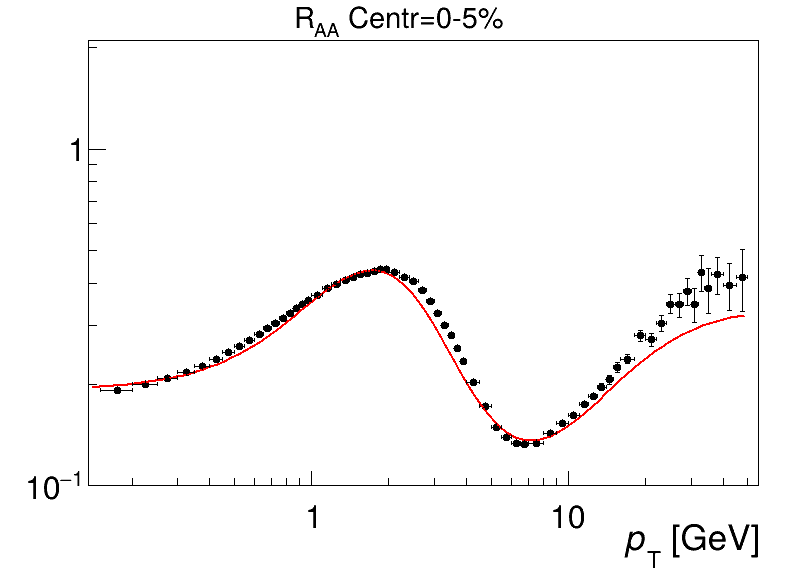}
  \includegraphics[width=0.3\linewidth]{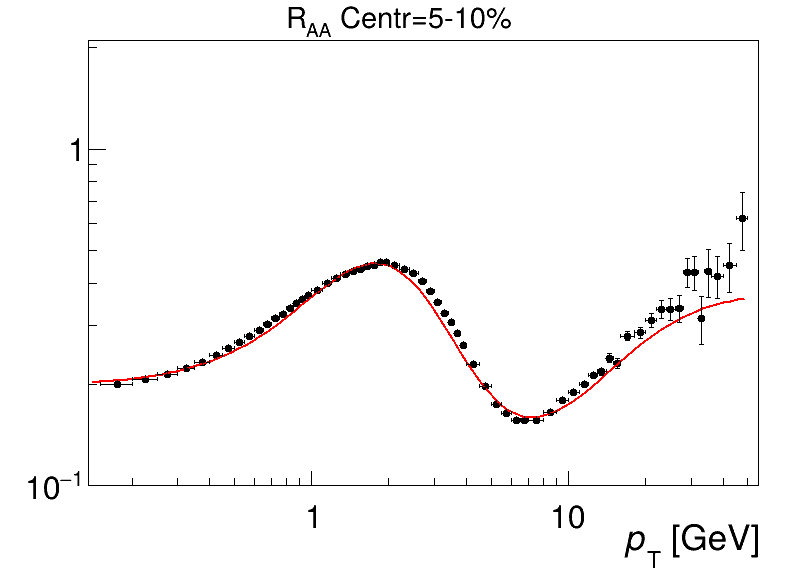}
  \includegraphics[width=0.3\linewidth]{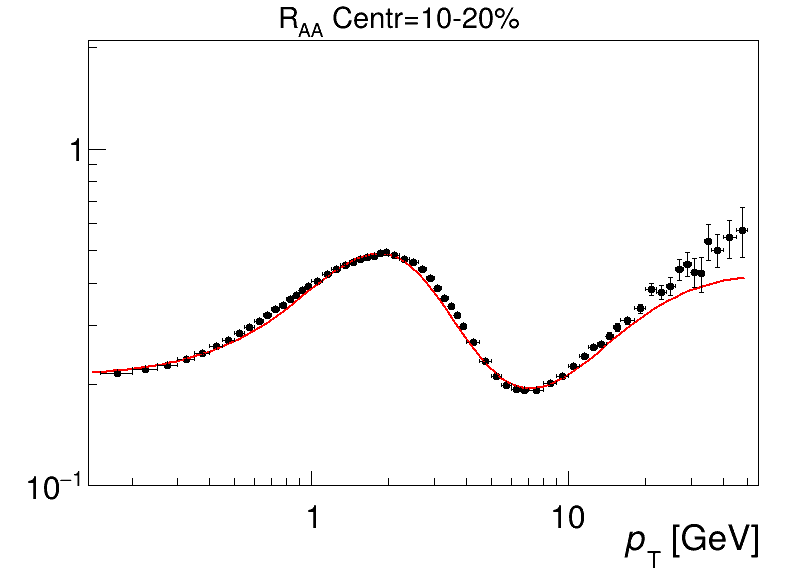}
  \includegraphics[width=0.3\linewidth]{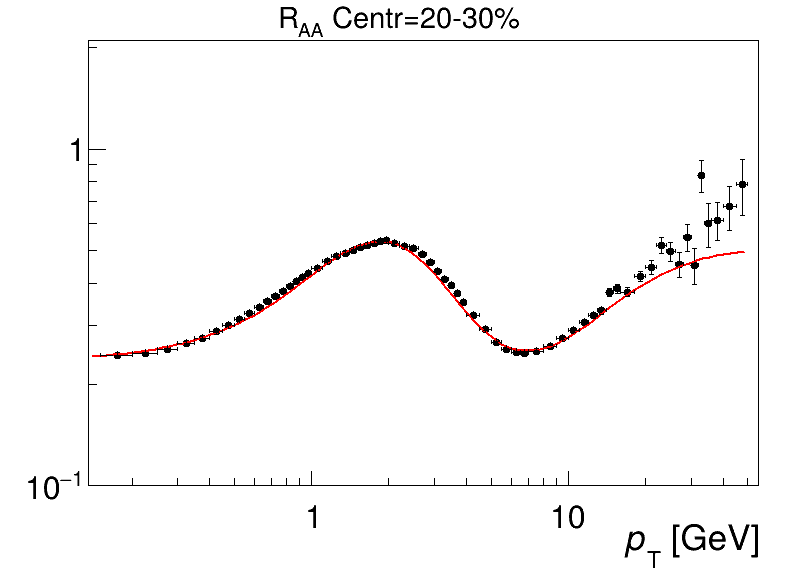}
  \includegraphics[width=0.3\linewidth]{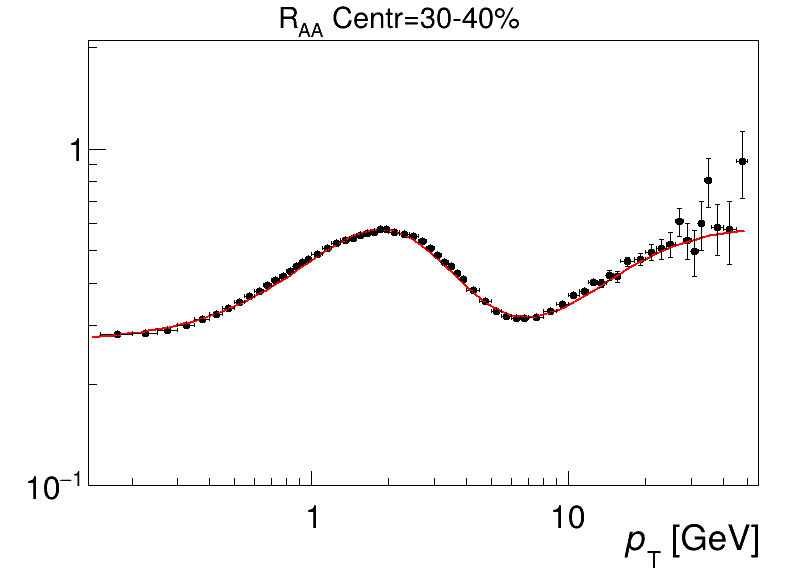}
  \includegraphics[width=0.3\linewidth]{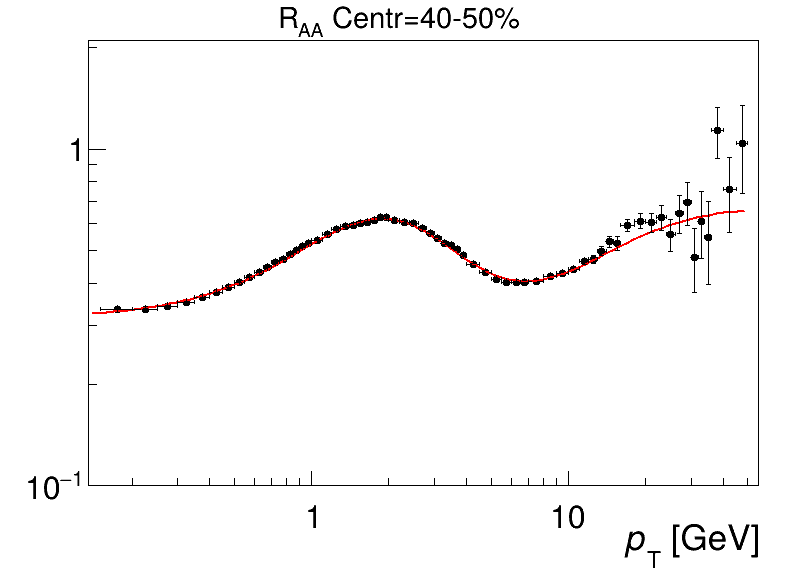}
  \includegraphics[width=0.3\linewidth]{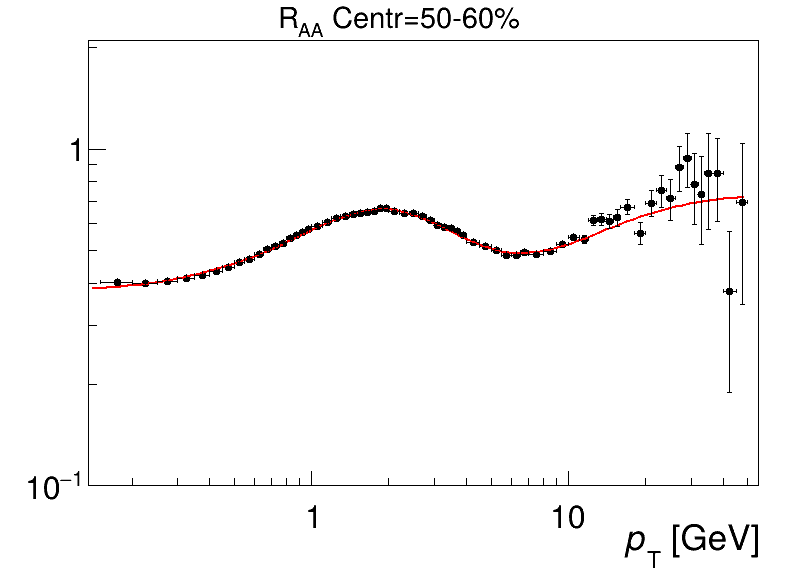}
  \includegraphics[width=0.3\linewidth]{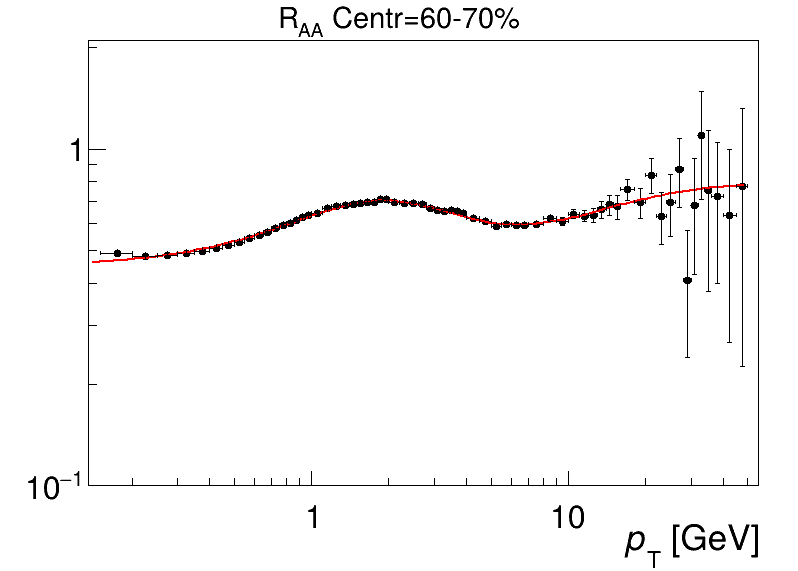}
  \includegraphics[width=0.3\linewidth]{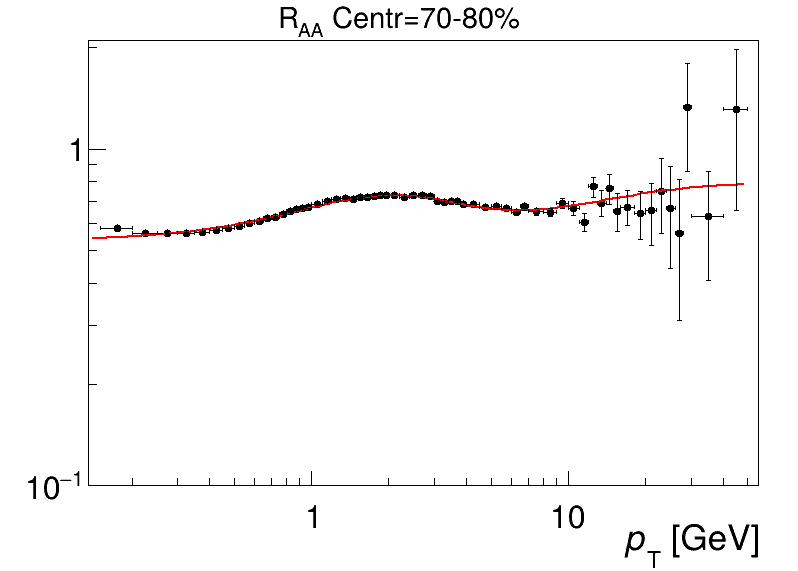}
  \caption{ALICE experimental data (black circles) for the $R_{AA}$ of charged particle production in PbPb collisions at $\sqrt{S_{NN}}=2.76$ TeV \cite{abelev2013centrality}, compared with theoretical results (full line). The plots are for centrality bins $0-5\%$,$5-10\%$, $10-20\%$, $20-30\%$, $30-40\%$, $40-50\%$, $50-60\%$ and $70-80\%$ respectively.}
  \label{fig:RAA276}
\end{figure}

\begin{figure}[H]
  \centering
  \includegraphics[width=0.3\linewidth]{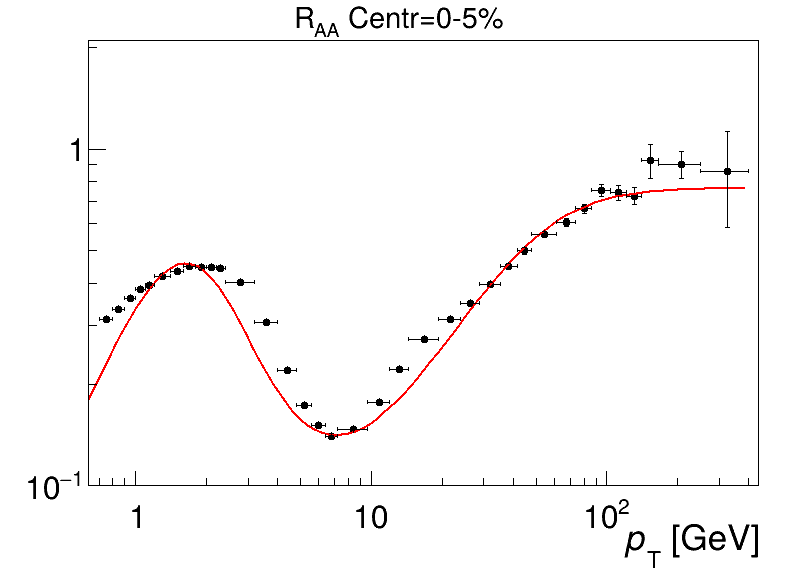}
  \includegraphics[width=0.3\linewidth]{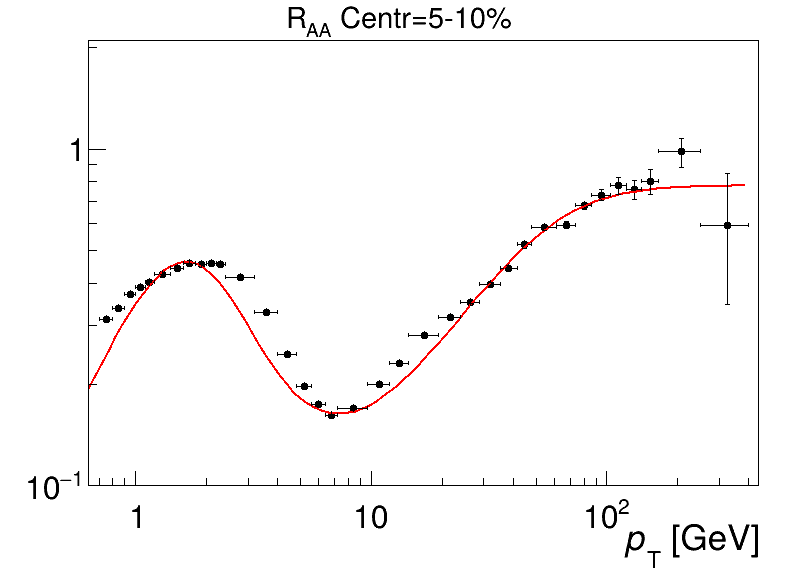}
  \includegraphics[width=0.3\linewidth]{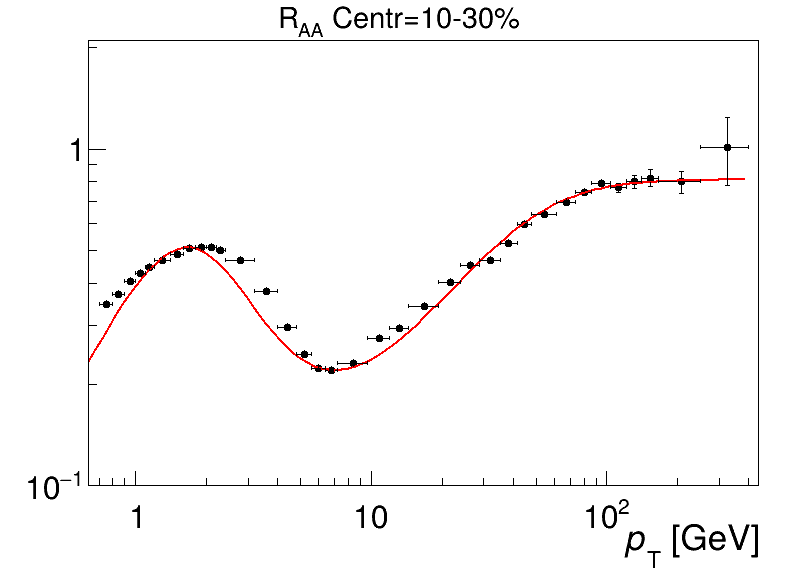}
  \includegraphics[width=0.3\linewidth]{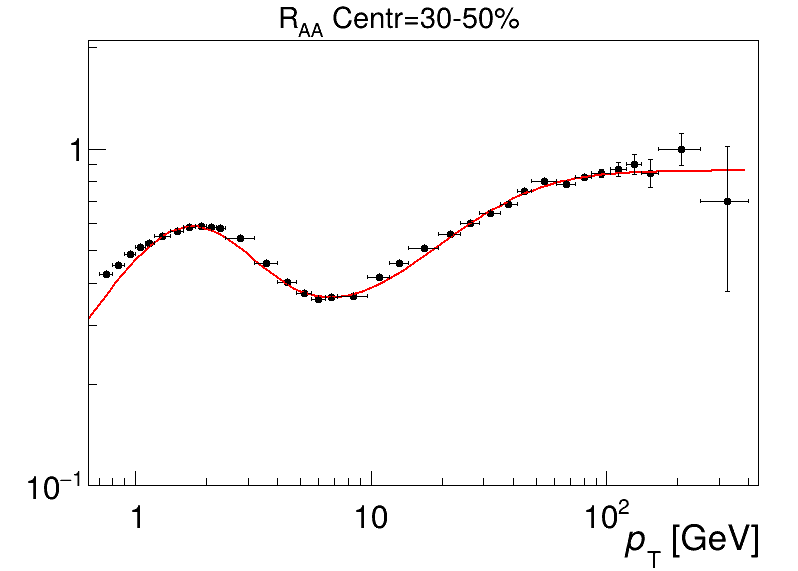}
  \includegraphics[width=0.3\linewidth]{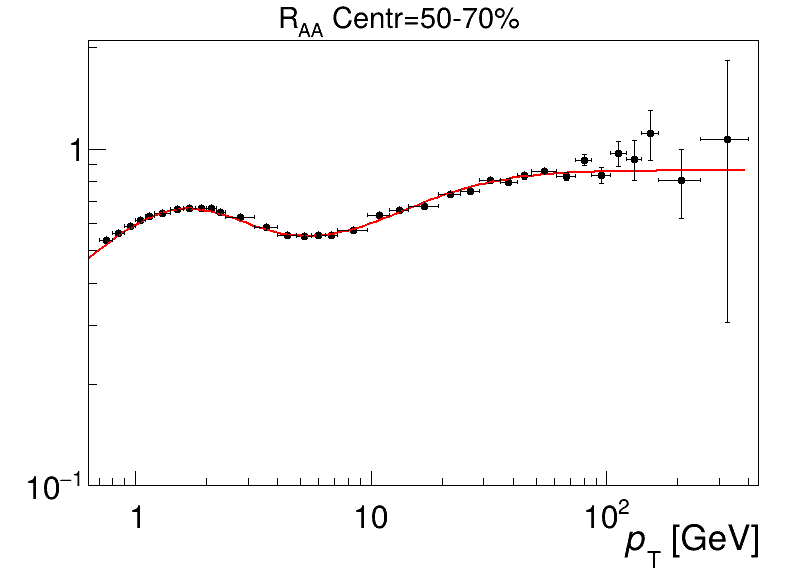}
  \includegraphics[width=0.3\linewidth]{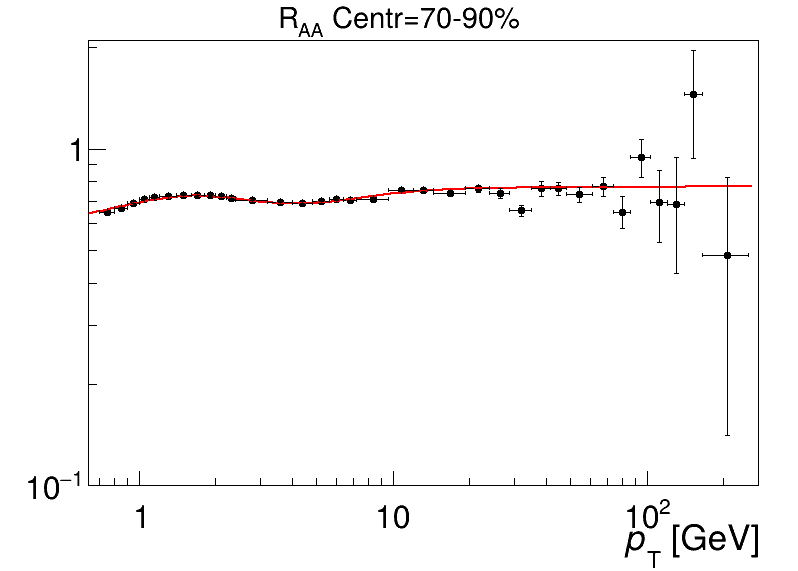}
  \caption{CMS experimental data (black circles) for $R_{AA}$ of charged particle production in PbPb collisions at $\sqrt{S_{NN}}=5.02$ TeV \cite{khachatryan2017charged} compared with theoretical results (full line). The plots are for centrality bins $0-5\%$,$5-10\%$, $10-30\%$, $30-50\%$, $50-70\%$ and $70-90\%$ respectively.}
  \label{fig:RAA502}
\end{figure}

Taking this into account, the nuclear modification factor can be expressed as
\begin{equation}
R_{AA} =
K N_{AA}^{-d} ~(1 - \epsilon)^{-d} 
\frac{V_{AA}}{V_{pp}} 
\frac{f\left(L_{-u}\left[L_u[p]\exp(A\tau)\right]\right)}{f_{\Delta p_l}(p_T)} \,,
\end{equation}
where $K$ is a constant and $V_{AA}$, $V_{pp}$ are the freeze-out volumes in $AA$ and $pp$ collisions, respectively (Figs.~\ref{fig:RAA276} and \ref{fig:RAA502}). Since $K$, $V_{AA}$, and $V_{pp}$ are not independently known, it is convenient to group them into a single parameter:
\begin{equation}
R_{AA}^0 = K N_{AA}^{-d} \frac{V_{AA}}{V_{pp}} \,.
\end{equation}





\subsection{Dynamical effects on the momentum}

\begin{figure}[H]
  \centering
    \includegraphics[width=0.32\linewidth]{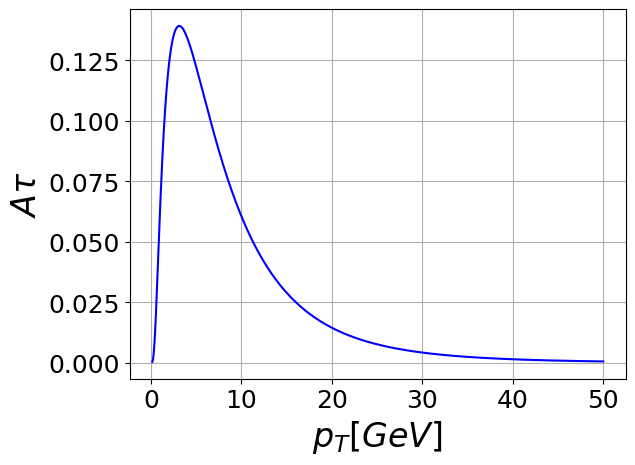}
    \includegraphics[width=0.32\linewidth]{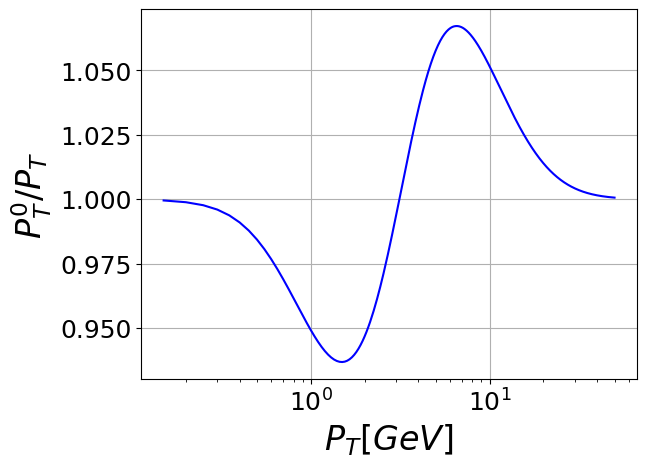}
    \includegraphics[width=0.32\linewidth]{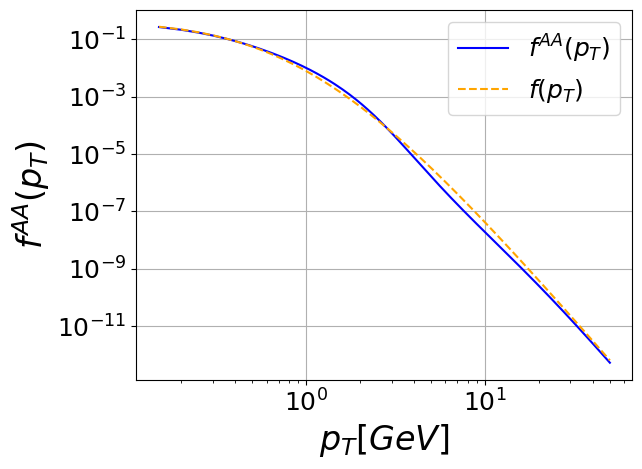}
\caption{{\it Left panel:} The drag coefficient as a function of the transverse momentum. {\it Middle panel:} The oscillation of the initial momentum $p_T^o$ of the parton as a function of the observed momentum $p_T$. {\it Right panel:} The $AA$ momentum distribution after dynamical effects, $f^{AA}$, as a function of the observed transverse momentum (solid line), compared to the $pp$ transverse momentum distribution for $pp$ collision, $f$.}
 \label{fig:transformedmomenta}
\end{figure}

Let us first consider only the drift term of the PPE. The behaviour of the drift coefficient, $A(p)$, as it relates to the parton's momentum in the fluid frame, is crucial for accurately describing the high transverse momentum sector of $R_{AA}$. While a detailed calculation of $A(p)$ is complex~\cite{Megas2023}, a phenomenological approach is sufficient for this study. The momentum dependence can be modelled by considering that the running coupling, which quantifies the strength of the partons' interaction with the medium, follows a $q$-exponential function with $q=1.16$~\cite{Deppman2020}. Therefore, the momentum-dependent drift coefficient is given by:
\begin{equation}
    A(p_T)= A_o \exp_q\left[-\bar{m}_T/T_D \right] \,, \label{eq:driftPt}
\end{equation}
where $T_D$ is associated with the energy scale of the dynamical effects (Fig.~\ref{fig:transformedmomenta}).


The diffusion effects, governed by the stochastic term in the PPE, were also evaluated (see Supplementary Material for additional information). However, the analysis shows that the diffusion effects in the nuclear modification factor are negligible. It is important to underline, at this point, that the initial momentum of the partons considered here is almost perpendicular to the collision axis. The dynamical effects alter the transverse momentum of the partons, as measured in the laboratory frame. Among these effects, the drag effect is the most relevant for the nuclear modification factor.


\section{Results and discussion}

%
%

The formalism developed above was applied to generate theoretical curves fitted to experimental data on charged particle production in PbPb collisions at $\sqrt{s_{NN}} = 2.76$ TeV and $\sqrt{s_{NN}} = 5.02$ TeV. The analysis focuses on the nuclear modification factor $R_{AA}$ as a function of the parton transverse momentum, measured for various collision centralities~\cite{abelev2013centrality, khachatryan2017charged}. The results are presented in Figs.~\ref{fig:RAA276} and~\ref{fig:RAA502} for the respective energies and centralities. For taking into account the effects of fluid expansion, the temperature of the $AA$ distribution, $T_{kin}$, is allowed to differ from that for the $pp$ collisions.

A fair agreement is observed between the theoretical predictions and the experimental data, indicating that the dynamical effects captured by the model can reproduce the oscillatory behaviour of the $R_{AA}$. Discrepancies appear in both the low- and high-$p_T$ regions (up to 50\% centrality), likely due to relativistic effects not included in the current formulation. It is worth noting that, after the Lorentz transformation, both low and high momenta in the center-of-mass frame correspond to relatively high momenta in the local fluid rest frame.

A central aspect of the observed agreement between data and calculations is the inclusion of the nonextensive aspects of the multiparticle production through Eq.~(\ref{eq:nonext}), which determine the overall intensity of the nuclear deformation. This represents strong evidence of the nonextensivity in the particle production. The fact that it is reproduced by employing the fractal exponent $d$ aligns with recent connections between Tsallis Statistics and fractal structures in QCD~\cite{Deppman2020}.

\begin{figure}[H]
  \centering
  \includegraphics[width=0.45\linewidth]{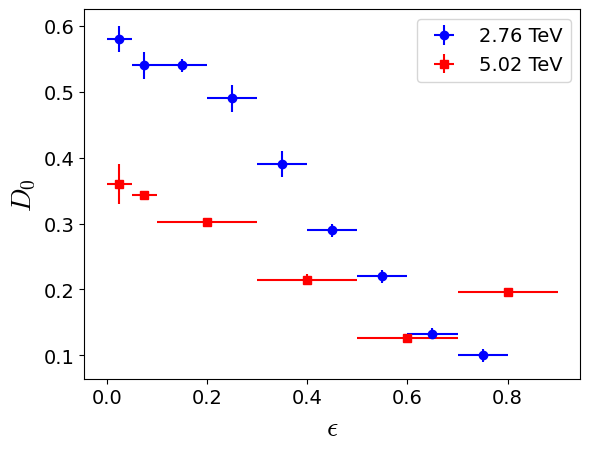}
  \includegraphics[width=0.45\linewidth]{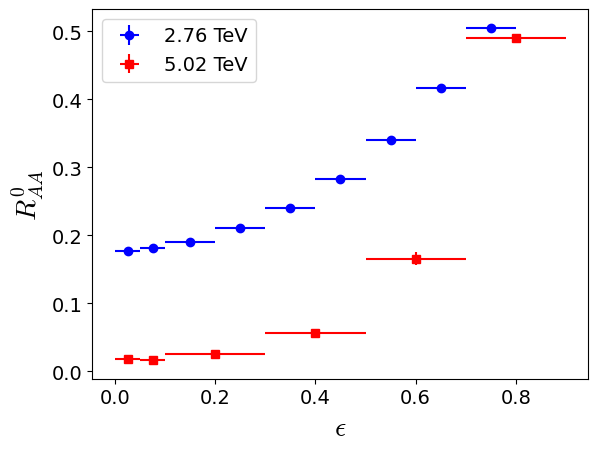}
  \includegraphics[width=0.45\linewidth]{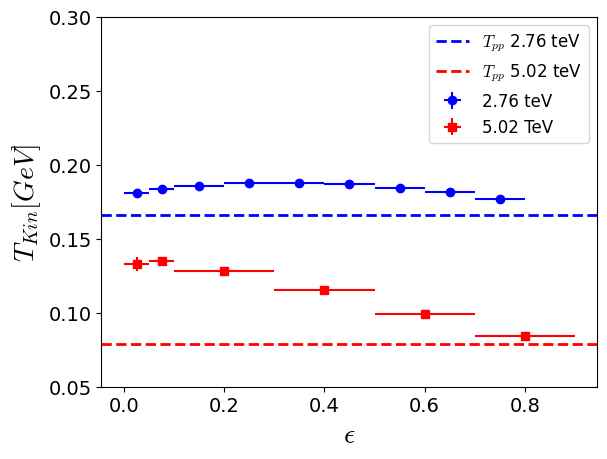}
  \includegraphics[width=0.45\linewidth]{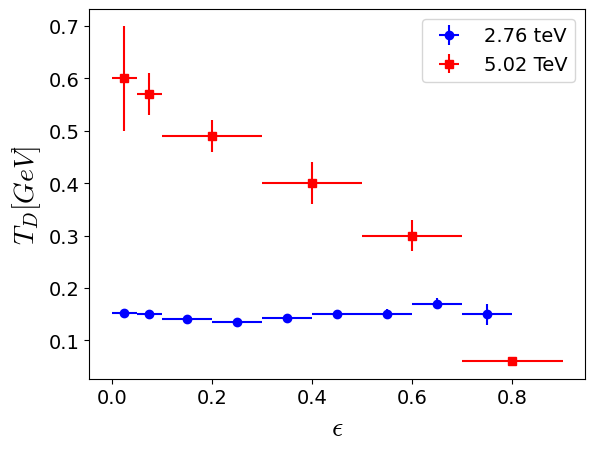}
 \caption{Fitted parameters as a function of centrality for $\sqrt{s_{NN}}=2.76$ TeV (blue) and $\sqrt{s_{NN}}=5.02$ TeV (red).}
  \label{fig:ParPlots}
\end{figure}

The free parameters used in the fits are $u$, $D_0=A_0 \tau$ (where $\tau$ is the freeze-out time), $R_{AA}^0$, $T_{Kin}$ (the temperature derived from the $AA$ spectra), and $T_D$ (the temperature associated with the drag function $A(p_T)$). The parameter $T_{pp}$ in the denominator of $R_{AA}$ is fixed from the description of $pp$ spectra: $T_{pp} = 0.166$ GeV for the 2.76 TeV data and $T_{pp} = 0.079$ GeV for the 5.02 TeV data, values that result from fitting to the $pp$-collisions data (see Supplementary Material for details). The dependencies of the parameters on the collision centrality, for both energies, are shown in Fig.~\ref{fig:ParPlots}.

The interplay between Lorentz transformations and drag effects establishes a relation between the original parton momentum $p_T^o$ and the observed momentum $p_T$, as shown in the Supplementary Material. The dynamics induce slight modifications to the momentum, resulting in an approximately linear relationship between $p_T^o$ and $p_T$. In this case, a log-periodic oscillation of the original momentum emerges, which gives rise to the oscillatory behaviour of $R_{AA}$ and establishes a connection with the complex-$q$ approach, as discussed below.

Figure~\ref{fig:ParPlots} presents the best-fit values for additional model parameters. The parameter $D_0$ (top-left) is associated with the phenomenological drift term, which depends on the particle momentum, as defined in Eqs.~(\ref{eq:DragEffect}) and~(\ref{eq:driftPt}). The plots show an approximately linear decrease of $D_0 $ with the centrality parameter $\epsilon$, a trend observed at both collision energies. Since $A_0$ is related to the drag effect and is primarily sensitive to the medium's properties rather than its geometry, it is reasonable to assume that $A_0$ remains nearly constant with centrality. 
Thus, the observed linear behaviour of the parameter $D_o=A_o \tau$ can be attributed to the freeze-out time, $\tau$, which varies accordingly with the centrality. As $\tau$ varies up to approximately $7$ fm/c~\cite{Harris2024}, it is reasonable to conclude that $A_o \cong 0.1$ c/fm. This indicates that the medium in central collisions has a longer lifetime, consistent with the expectation that central collisions produce a more complex and long-lived system.

The top-right panel shows the parameter $R_{AA}^0$, which remains roughly constant in central collisions but exhibits a slight increase toward more peripheral ones. This trend can be attributed to geometric effects, particularly the ratio $V_{AA} / V_{pp}$, which becomes less reliably described by a linear dependence on centrality in more peripheral events. Although the trend is similar at both energies, the absolute values of $R_{AA}^0$ differ significantly.

The bottom-left panel displays the temperature $T_{AA}$ of the system formed in $AA$ collisions, which is treated independently from the temperature in $pp$ collisions. At 2.76~TeV, the temperature is nearly constant and slightly exceeds the $pp$ value. At 5.02~TeV, however, a clear centrality dependence emerges. Since higher energies involve larger parton momenta in the medium formation, this behavior may result from the neglect of relativistic effects in the PPE. Therefore, the observed centrality dependence likely stems more from approximations in the analytical treatment than from intrinsic physical properties of the medium's evolution.

The bottom-right panel presents the parameter $T_D$, which sets the energy scale in the phenomenological drag coefficient. At 2.76~TeV, it shows little dependence on centrality. At 5.02~TeV, larger uncertainties prevent a definitive conclusion, though the average value is slightly higher. Given the non-relativistic nature of the approximation, we refrain from assigning strong physical meaning to these results, including those for $T_{Kin}$. Nonetheless, the smooth behaviour of $T_D$ supports the robustness of the phenomenological drag description and suggests it may serve as a reliable starting point for more realistic modelling.

The results indicate that the radial expansion velocity of the fluid is essentially independent of collision centrality, with values of $u = 0.99899 \, c$ at 2.76 TeV and $u = 0.9985 \, c$ at 5.02 TeV. This shows an almost negligible dependence on both centrality and collision energy. These values are, however, significantly higher than those predicted by hydrodynamical models, which typically yield fluid velocities between 50\% and 80\% of the speed of light. From a phenomenological perspective, such fluid velocities appear unrealistic.

In the blast-wave approach, the fluid velocity is constrained to remain above the speed of sound, which can reach up to $c_s \cong 0.6$ for the QGP~\cite{Borsnyi2010,Bazavov2014}. In contrast, hydrodynamical models—where the fluid expansion is governed by pressure waves—restrict the velocity to remain below the speed of sound. Furthermore, the simplifications adopted in the present calculations, such as modelling freeze-out through a narrow cylindrical shell containing the fluid, may artificially increase the velocity due to the higher energy density involved in the expansion. Another important factor is the assumption that the QGP behaves as a pionic fluid, with the mass of the fluid quasi-particle equal to the pion mass. If instead the effective mass is taken as $m \cong 2.7$~GeV, the fluid velocity decreases to $u=0.8$.

Beyond these physical constraints, the analytical framework employed here relies on several idealisations aimed at establishing a manageable connection with the complex-$q$ formalism. The assumption of a freeze-out surface concentrated in a narrow cylindrical shell artificially increases the energy density and, consequently, the velocity. Similarly, the adoption of a constant fluid velocity, independent of the radial coordinate $r$, also contributes to an overestimation. These assumptions can be relaxed in future numerical implementations. Accordingly, the present work should be regarded as a first attempt to describe the nuclear modification factor as a dynamical effect, with substantial refinements still required in the theoretical framework.

Since the dynamical approach results can describe the nuclear modification factor, and since it is based on analytical expressions, it is possible to link the present approach with the complex-$q$ approach. The log-periodic oscillation of $p_T^o/p_T$ opens the opportunity to link the dynamics effects with the log-periodic oscillation observed in~\cite{Wilk2014,Rybczyski2015}. This section establishes a possible link between the complex-$q$ approach and the dynamical approach, aiming to express the real and imaginary parts of $q$ in terms of the dynamical parameters.

The starting point is the transformation between $p_T$ and $p^o_T$, given by Eq.~(\ref{eq:popT}). Using the Lorentz transformation, that equation results in~\footnote{In Eq.~(\ref{eq:popT1}), $v$ and $v_T$ denote the components of the 3-vectors in the transverse direction.}
  \begin{equation}
   \frac{p_T^o}{p_T} = \frac{p_M}{p_T} \left(  \frac{X}{v} + \sqrt{1 + X^2} \right) \,, \quad \textrm{where} \quad X \equiv  \gamma_{v} \, \gamma_{v_T} \left(  v_T - v \right) \exp\left[A(\bar p_T)\tau \right]   \,,  \label{eq:popT1}
  \end{equation}
with $v = p_M / (p_M^2 + m^2)^{1/2}$, $v_T = p_T/ m_T$ and $\gamma_r = (1-r^2)^{-1/2}$ for $r = v, v_T$.

The results obtained in the previous sections show that the ratio $p_T^o/p_T$ remains close to unity in the relevant range of $p_T$. A Taylor expansion around $A(\bar p_T) \tau = 0$ leads to
  \begin{equation}
     \frac{p_T^o}{p_T} \simeq 1 + \frac{p_T^2 - p_M^2}{p_T^2 + p_M^2}  A(\bar p_T) \tau   \simeq 1 + A_0 \tau   \frac{\left(p_T^2 - p_M^2\right)}{p_T^2 + p_M^2}  \exp_q\left[ \frac{a}{T}-  \frac{\left(p_T^2+ p_M^2\right)}{2 p_T p_M} \frac{m}{T}  \right]   \,,  \label{eq:popT3_main}
  \end{equation}
  where in the second equality we have considered an expansion of $A(\bar p_T)$ at small $m$. In this equation $a$ is a constant that in the following we will set to zero for simplicity. Defining the variable $y \equiv \bar m_T / m$ and performing a Taylor expansion around $y=1$, which corresponds to $\bar p_T = 0$,  Eq.~(\ref{eq:popT3_main}) results in
  \begin{equation}
  \frac{p_T^o}{p_T} \simeq 1 + 2 A_0 \tau m \, (y-1) \exp_q\left[ - y \, m/T \right]   \,. \label{eq:popT4}
  \end{equation}
Using now that $\sin(y-1) \simeq  y-1 \simeq \sin(\log y)$, one arrives at the following result
 \begin{equation}
  \frac{p_T^o}{p_T} \simeq 1 + 2 \tilde A_0 \tau \, (1+z)^{-m_0} \cos\left[ c_1 \log(1+z) \right]   \,,  \label{eq:popT51}
  \end{equation}
where 
\begin{equation}
1 + z = e^{-\frac{\pi}{2 c_1}} \frac{1+x}{1+x_o}  \,, \qquad y = x/x_o \,, \quad x_o = (q-1) m / T \,, \quad c_1 = 1 + 1/x_o \,,
\end{equation}
while $\tilde A_0 =\left( e^{\frac{\pi}{2c_1}} (1+x_o)\right)^{\frac{1}{1-q}}  m  \, A_0$, and $m_0 = 1/(q-1)$. This corresponds to a complex-$q$ Tsallis distribution of the form
\begin{equation}
\frac{p_T^o}{p_T} - 1 \simeq (1+z)^{-\textrm{Re}(m_k)} \sum_{k=0}^\infty w_k \cos\left[ \textrm{Im}(m_k)  \log(1+z)\right] \,,  \label{eq:popT52}
\end{equation}
with
\begin{equation}
m_k = \frac{1}{q-1} + i k \left( 1 + \frac{1}{x_o}\right)  \,,\qquad (k = 0, 1, 2, \cdots) \,.  
 \end{equation}
Only the term $k=1$ is present in Eq.~(\ref{eq:popT52}), i.e. $w_1 = 2 \tilde A_0 \tau \ne 0$ and $w_n = 0 \; \forall n \ne 1$. In an expansion around $q=1$, one has
\begin{equation}
q_k = 1 + \frac{1}{m_k} \simeq 1 + \frac{ 1 - i \xi_k}{1 + \xi_k^2} (q-1) + O\left((q-1)^2 \right)\,, \label{eq:qk_expansion}
\end{equation}
where $\xi_k \equiv kT/m$. Notice that $q_k \simeq q$ for $T/m \to 0$ if $q$ is close to 1.

\section{Conclusions and outlook}
In summary, this work calculates the nuclear suppression factor of charged particles in Pb–Pb collisions at two collision energies: 2.76 TeV and 5.02 TeV. The novel approach incorporates dynamical effects within the framework of the Plastino–Plastino Equation, considering a $q$-exponential ansatz for the drag coefficient, which leads to momentum depletion. The effect of diffusion is found to be negligible in our calculations.

The results indicate that the oscillations observed in the $R_{\text{AA}}$ data have a dynamical origin, stemming from the ratio between the initial and depleted momenta in the center-of-mass frame. This oscillatory behavior gives rise to a complex $q$-parameter. The theoretical model shows good agreement with experimental data at $\sqrt{s} = 2.76$ TeV across all centralities and momentum ranges. Minor deviations in the high-momentum region can be attributed to simplifications introduced to obtain analytical expressions for the dynamical effects.

The satisfactory agreement between the model and the data underscores the relevance of log-periodic oscillations, which have been observed in other analyses as well. The connection established between the complex-$q$ formalism and the dynamical approach provides a new interpretation of complex-$q$ parameters, complementing existing interpretations.

This work can be extended to improve accuracy by allowing for numerical solutions of the fully relativistic Plastino–Plastino Equation. Such an approach would enhance our understanding of the role played by the drag transport coefficient in parton dynamics within the medium. 

\section*{Acknowledgements}
In preparation for this publication, we used the resources of the
Centre for Computation and Computer Modelling of the Faculty of Exact
and Natural Sciences of the Jan Kochanowski University in Kielce,
modernised from the funds of the Polish Ministry of Science and Higher
Education in the “Regional Excellence Initiative” programme under the
project RID/SP/00015/2024/01. G. Wilk was supported in part by the
Polish Ministry of Education and Science, Grant No. 2022/WK/01.
R. Baptista's work was supported by the project
RID/SP/00015/2024/01. T. Bhattacharyya acknowledges funding from the
European Union's HORIZON EUROPE programme, via the ERA Fellowship
Grant Agreement number 101130816. A. Deppman is partially supported by
the CNPq, grant 306093/2022-7, and by FAPESP grant
24/01533-7. E. Meg\'{\i}as is supported by the ``Proyectos de
Investigaci\'on Precompetitivos'' Program of the Plan Propio de
Investigaci\'on of the University of Granada under grant PP2025PP-18,
and by the Junta de Andaluc\'{\i}a under grant~FQM-225.


\section{Supplementary Material}
\label{sec:Suppl_Material}

\subsection{Fittings of $pp$ collisions}

\begin{figure}[H]
  \centering
  \includegraphics[width=0.45\linewidth]{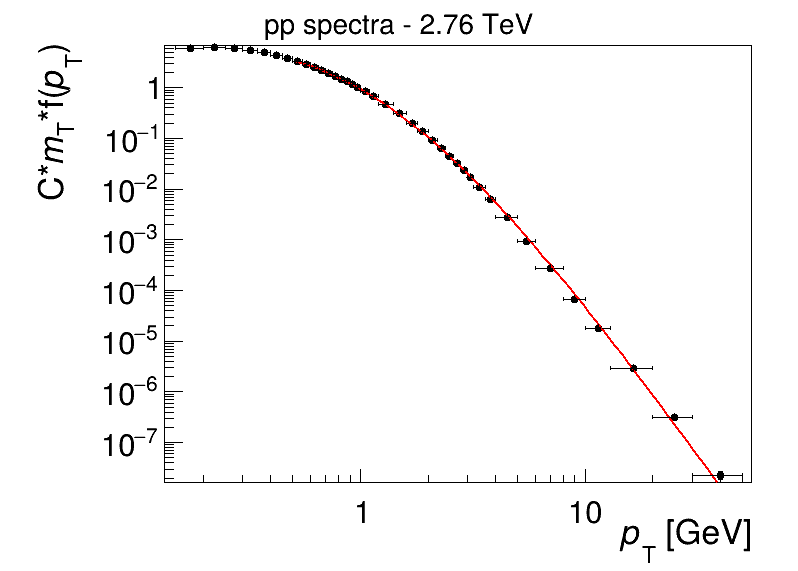}  \hspace{0.5cm}
  \includegraphics[width=0.45\linewidth]{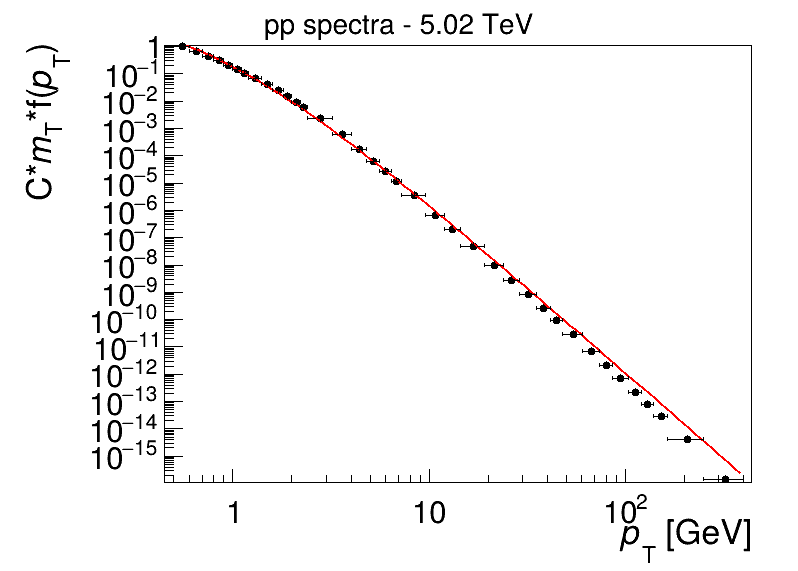}
 \caption{Fittings of the transverse momentum spectra for $pp$ data for $2.76$~GeV (left) and $5.02$~GeV (right) collsion energies. $C$ is a fitting constant, $m_T$ the transverse mass and $f(p_T$) the momentum distribution function.}
  \label{fig:pt0pt}
\end{figure}  




\subsection{Connection with wounded nucleon model}

The collision energy is proportional to the number of participant nucleons, $N_{\text{part}}$. We have checked using the Glauber model-based GLISSANDO Monte-Carlo code \cite{Bozek:2019wyr} for the Pb-Pb collisions at 2.76 TeV that the number of participant nucleons can be written as,
%
%
\begin{equation}
    N_{\text{part}}=5.9+402 \left[1- \epsilon \right]^{2.71},
\end{equation}
where $\epsilon$ is the centrality. Thus, the collision energy can be expressed as $E \propto N_{\text{part}}$, leading to
\begin{equation}
    \frac{N}{E} \propto \left[ 5.9+402 \left[1- \epsilon \right]^{2.71} \right]^{-d}.
\end{equation}


Using the same numerical framework, we also find that
$N_{\text{coll}} \approx 0.35 ~[N_{\text{part}}]^{1.4}$.

\subsection{Inclusion of the diffusive effects}

%

The effects of the diffusion on the transverse momentum distribution are characterized by the broadening of the $ p_T$ distribution as the parton travels through the medium. The width of the distribution is given by~\cite{Megias_2024}
\begin{equation}
    \sigma(t)=\sigma_o \left[(1-\kappa) \exp(-\xi At)+\kappa \right]^{1/\xi}  \,,
\end{equation}
where $\sigma_o$ is the initial distribution width, $\kappa$ is a constant that depends on the transport coefficients $A$ and $D$, and $\xi=5-3q$.

The effects of the inclusion of the diffusion effects in $R_{AA}(p_T)$ were included by considering that the original transverse momentum of the parton, before its interaction with the medium, not given by Eq.~(\ref{eq:popT}) anymore, but by a set of value ${p_m}$ which, after the dynamics effects, will end up being $p_T$ with some probability given by the $q$-gaussian probability distribution
\begin{equation}
    G_q(p_m,p)=  e_q\left( -\frac{(p-p_m)^2}{2 \pi \sigma(\tau)^2}\right)   \,,
\end{equation}
The nuclear-nuclear distribution becomes
\begin{equation}
    f_D^{AA}(p_T)= {\mathcal N} \int G_q\left(L_u[p_m],L_u[p_T]\right) f( L_{-u}\left[L_u[p_m] \exp(A \tau)\right]) ~dp_o \,, \label{eq: fAA-dyn}
\end{equation}
with
\begin{equation}
    {\mathcal N}^{-1}=\int G_q\left(L_u[p_m],L_u[p_T]\right)~dp_o  \,,
\end{equation}
where ${\mathcal N}$ is a normalization constant. 

\subsubsection{Dynamical effects on the original and observed momenta}

\begin{figure}[H]
  \centering
  \includegraphics[width=0.45\linewidth]{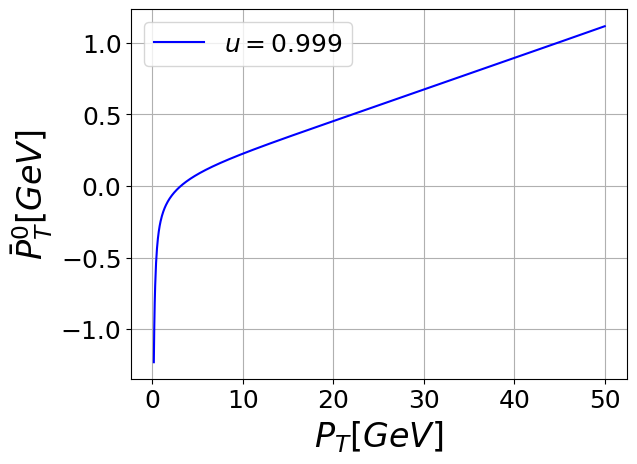} \hspace{0.5cm}
  \includegraphics[width=0.45\linewidth]{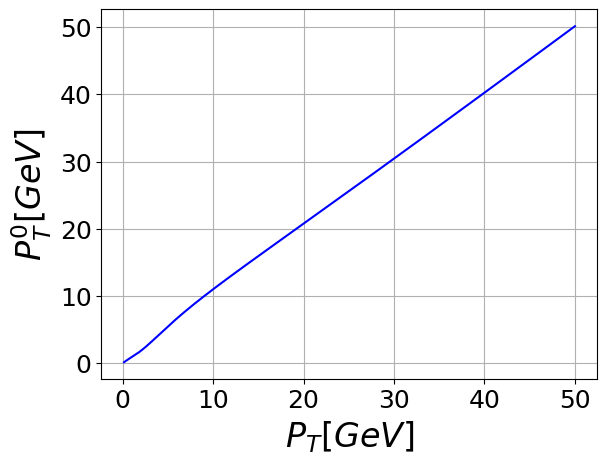}    
 \caption{{\it Left panel:} The initial parton momentum in the local rest frame of the fluid. {\it Right panel:} The momentum of the parton in the moment it was created, $p_T^o$, as a function of the observed momentum $p_T$. }
 \label{fig:transfmomenta}
\end{figure} 

\subsection{\bf{Log-oscillations in $R_{AA}$}}

\begin{figure}[H]
  \centering
  \includegraphics[width=0.45\linewidth]{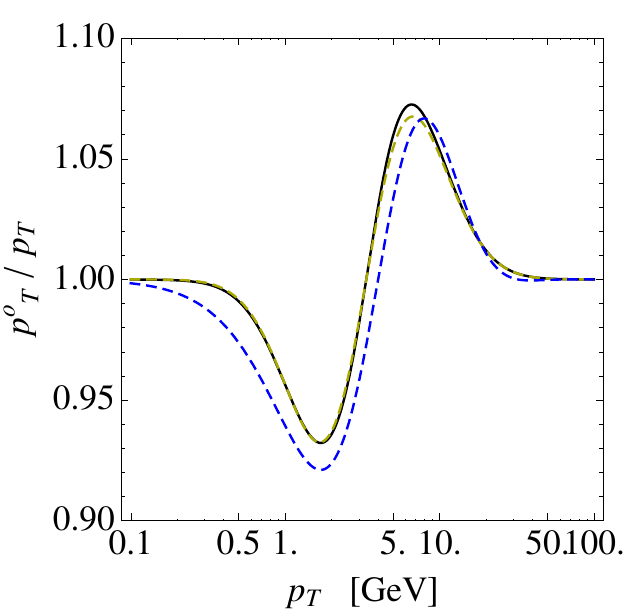}  \hspace{0.5cm}
\includegraphics[width=0.43\linewidth]{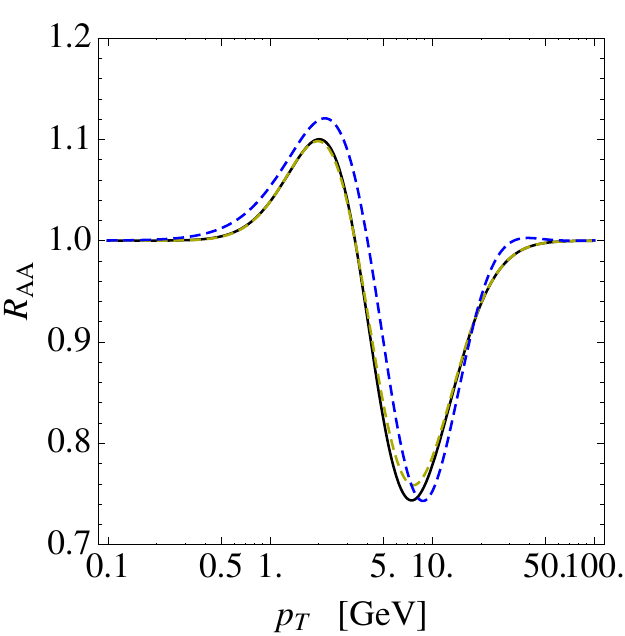}
 \caption{{\it Left panel:} Ratio $p_T^o/p_T$ as a function of $p_T$. We display in solid (black) the result given by Eq.~(\ref{eq:popT1}), in dashed (yellow) the approximation given by Eq.~(\ref{eq:popT3_main}), and in dashed (blue) the approximation given by Eq.~(\ref{eq:popT4}). {\it Right panel:} $R_{AA}$ as a function of $p_T$. We display the results of the left panel transformed by using Eq.~(\ref{eq:RAA_approx}). The lines follow the same pattern as in the left panel.}
  \label{fig:RAA}
\end{figure}

Let us study how the log-oscillations in $p_T^o/p_T$ produce log-oscillations in $R_{AA}$. The particle momentum distribution is
\begin{equation}
f^{AA}(p_T) = f(p_T^o) = \left[ 1 + (q-1) \frac{p_T^o}{\lambda}\right]^{-\frac{1}{q-1}} = \left[ 1 + (q-1) \frac{p_T}{\lambda} \frac{p_T^o}{p_T}\right]^{-\frac{1}{q-1}}  \,.
\end{equation}
By using now that $p_T^o/p_T \simeq 1 + Y(p_T)$ where $Y(p_T)$ is a small oscillatory contribution, one can make a Taylor expansion of this expression up to order $O(Y)$, to get
\begin{equation}
f^{AA}(p_T) \simeq \left[ 1 + (q-1) \frac{p_T}{\lambda}\right]^{-\frac{1}{q-1}} \times  \left[ 1 - \frac{p_T/\lambda}{1 + (q-1) p_T/\lambda} Y(p_T) \right] \,.
\end{equation}
Moreover
\begin{eqnarray}
R_{AA}(p_T) &=& \left[ N_{AA} (1-\epsilon) \right]^{-d}\frac{V_{AA}}{V_{pp}} \frac{f^{AA}(p_T)}{f(p_T)}    \nonumber \\
&\simeq& \left[ N_{AA} (1-\epsilon) \right]^{-d}\frac{V_{AA}}{V_{pp}} \left[ 1 - \frac{p_T/\lambda}{1 + (q-1) p_T/\lambda} Y(p_T) \right]  \,.  \label{eq:RAA_approx}
\end{eqnarray}
Then, the oscillations in $Y(p_T)$ induce oscillations in $f^{AA}(p_T)$ and $R_{AA}(p_T)$. The results for $p_T^o/p_T$  and $R_{AA}$ as a function of $p_T$ are displayed in Fig.~\ref{fig:RAA}.

Other way to arrive at this result is by using
\begin{equation}
\left[1+(q-1)\frac{p_T^o}{\lambda}\right]^{\frac{-1}{q-1}} \cong \left[1+(q-1)\frac{p_T}{\lambda}\right]^{-\frac{1}{q-1}} \left(\frac{p_T^o}{p_T}\right)^{-\frac{1}{q-1}}\,,
~\end{equation}
which is valid for $p_T, p_T^o \gg \lambda$. Then
\begin{equation}
R_{AA}(p_T) \simeq \left[ N_{AA} (1-\epsilon) \right]^{-d}\frac{V_{AA}}{V_{pp}} \left(\frac{p_T^o}{p_T}\right)^{-\frac{1}{q-1}}  \simeq \left[ N_{AA} (1-\epsilon) \right]^{-d}\frac{V_{AA}}{V_{pp}} \left( 1 - \frac{1}{q-1} Y(p_T) \right)  \,.  \label{eq:RAA_approx2}
\end{equation}
This result is in agreement with Eq.~(\ref{eq:RAA_approx}) for $p_T \gg \lambda$, but not for $p_T \lesssim \lambda$. 
%
%
According to the experimental data for $R_{AA}(p_T)$, the corrections are of order $Y(p_T) \sim 10\%$ and $\frac{1}{q-1} Y(p_T) \sim 65\%$. 
Finally, an expression for $R_{AA}(p_T)$ valid even if $Y(p_T)$ is not small is 
\begin{eqnarray}
R_{AA}(p_T) &\simeq& \left[ N_{AA} (1-\epsilon) \right]^{-d}\frac{V_{AA}}{V_{pp}} \left(\frac{p_T^o}{p_T}\right)^{-\frac{1}{q-1}} = \left[ N_{AA} (1-\epsilon) \right]^{-d}\frac{V_{AA}}{V_{pp}} \left( 1 + \frac{q-1}{q-1} Y(p_T )\right)^{-\frac{1}{q-1}}  \nonumber \\
&=& \left[ N_{AA} (1-\epsilon) \right]^{-d}\frac{V_{AA}}{V_{pp}} \exp_q\left[ - \frac{1}{q-1} Y(p_T) \right]  \,.  \label{eq:RAA_approx3}
\end{eqnarray}

\bibliographystyle{ieeetr}
\bibliography{Manuscript}

@article{Alice2018,
  title = {Transverse momentum spectra and nuclear modification factors of charged particles in {pp},  {p-Pb} and {Pb-Pb} collisions at the {LHC}},
  volume = {2018},
  DOI = {10.1007/jhep11(2018)013},
  journal = {Journal of High Energy Physics},
  pages={013},
  author = {{Alice Collab.}},
  year = {2018}
}

@article{CMS2018,
  title = {Nuclear modification factor of {D}$^0$ mesons in {PbPb} collisions at $\sqrt{s_{NN}} = $ 5.02 {TeV}},
  volume = {782},
  DOI = {10.1016/j.physletb.2018.05.074},
  journal = {Physics Letters B},
  publisher = {Elsevier BV},
  author = {{CMS Collab.}},
  year = {2018},  month = jul,
  pages = {474–496}
}

@article{Bhattacharyya2024,
  title = {Jet quenching of the heavy quarks in the quark-gluon plasma and the nonadditive statistics},
  volume = {856},
  DOI = {10.1016/j.physletb.2024.138907},
  journal = {Physics Letters B},
  publisher = {Elsevier BV},
  author = {Bhattacharyya,  Trambak and Megías,  Eugenio and Deppman,  Airton},
  year = {2024},
  pages = {138907}
}

@article{Bhattacharyya2023,
  title = {Nonextensive {B}oltzmann transport equation: The relaxation time approximation and beyond},
  volume = {624},
  DOI = {10.1016/j.physa.2023.128910},
  journal = {Physica A: Statistical Mechanics and its Applications},
  publisher = {Elsevier BV},
  author = {Bhattacharyya,  Trambak},
  year = {2023},
  pages = {128910}
}

@article{Cooper1974,
  title = {Single-particle distribution in the hydrodynamic and statistical thermodynamic models of multiparticle production},
  volume = {10},
  DOI = {10.1103/physrevd.10.186},
  number = {1},
  journal = {Physical Review D},
  publisher = {American Physical Society (APS)},
  author = {Cooper,  Fred and Frye,  Graham},
  year = {1974},
  pages = {186–189}
}

@article{Plastino1995,
  title = {Non-extensive statistical mechanics and generalized {F}okker-{P}lanck equation},
  volume = {222},
  DOI = {10.1016/0378-4371(95)00211-1},
  number = {1–4},
  journal = {Physica A: Statistical Mechanics and its Applications},
  publisher = {Elsevier BV},
  author = {Plastino,  A.~R. and Plastino,  A.},
  year = {1995},
  pages = {347–354}
}

@article{Rocha2022,
  title = {Nonextensive Statistics in High Energy Collisions},
  volume = {4},
  DOI = {10.3390/physics4020044},
  number = {2},
  journal = {Physics},
  publisher = {MDPI AG},
  author = {Rocha,  Lucas Q. and Megías,  Eugenio and Trevisan,  Luis A. and Olimov,  Khusniddin K. and Liu,  Fuhu and Deppman,  Airton},
  year = {2022},
  pages = {659–671}
}

@article{Grigoryan2017,
  title = {Using the {T}sallis distribution for hadron spectra in pp collisions: Pions and quarkonia at $\sqrt{s} = 5-13000 \, \textrm{GeV}$},
  volume = {95},
  DOI = {10.1103/physrevd.95.056021},
  journal = {Physical Review D},
  publisher = {American Physical Society (APS)},
  author = {Grigoryan,  Smbat},
  year = {2017},
  pages={056021}
}

@article{Deppman2020,
  title = {Fractals,  nonextensive statistics,  and {QCD}},
  volume = {101},
  DOI = {10.1103/physrevd.101.034019},
  number = {3},
  journal = {Physical Review D},
  publisher = {American Physical Society (APS)},
  author = {Deppman,  Airton and Megías,  Eugenio and Menezes,  Debora P.},
  year = {2020},
  pages={034019}
}

@article{Megas2023,
  title = {Comparative study of the heavy-quark dynamics with the {F}okker-{P}lanck equation and the {P}lastino-{P}lastino equation},
  volume = {845},
  DOI = {10.1016/j.physletb.2023.138136},
  journal = {Physics Letters B},
  publisher = {Elsevier BV},
  author = {Megías,  Eugenio and Deppman,  Airton and Pasechnik,  Roman and Tsallis,  Constantino},
  year = {2023},
  pages = {138136}
}

@article{Megias_2024, 
title={Dynamics in fractal spaces}, volume={848},
journal={Physics Letters B},
author={Meg\'{\i}as, Eugenio and Khalili Golmankhaneh, Alireza and Deppman, Airton}, year={2024}
, pages={138370} }

@article{Wilk2014,
  title ={Tsallis distribution with complex nonextensivity parameter $q$},
  volume ={413},
  journal ={Physica A: Statistical Mechanics and its Applications},
  author ={Wilk,  G. and Włodarczyk,  Z.},
  year ={2014},
  pages ={53–58}
}

@article{Rybczyski2015,
  title = {System size dependence of the log-periodic oscillations of transverse momentum spectra},
  volume = {90},
  journal = {EPJ Web of Conferences},
  author = {Rybczyński,  Maciej and Wilk,  Grzegorz and Włodarczyk,  Zbigniew},
  editor = {Fabbri,  F. and Giacomelli,  P.},
  year = {2015},
  pages = {01002}
}

@article{PhysRevLett.84.31,
  title = {Equilibrium Distribution of Heavy Quarks in {F}okker-{P}lanck Dynamics},
  author = {Walton, D. Brian and Rafelski, Johann},
  journal = {Phys. Rev. Lett.},
  volume = {84},
  issue = {1},
  pages = {31--34},
  numpages = {0},
  year = {2000},
  month = {Jan},
  publisher = {American Physical Society},
  doi = {10.1103/PhysRevLett.84.31},
  url = {https://link.aps.org/doi/10.1103/PhysRevLett.84.31}
}

@article{abelev2013centrality,
  title={Centrality dependence of charged particle production at large transverse momentum in {P}b--{P}b collisions at $\sqrt{s}_{NN}$= 2.76 {TeV}},
  author={Abelev, Betty and Adam, Jaroslav and Adamov{\'a}, Dagmar and Adare, Andrew Marshall and Aggarwal, Madan Mohan and Rinella, G Aglieri and Agocs, Andras Gabor and Agostinelli, Andrea and Salazar, S Aguilar and Ahammed, Zubayer and others},
  journal={Physics Letters B},
  volume={720},
  number={1-3},
  pages={52--62},
  year={2013}
}

@article{khachatryan2017charged,
    author = "Khachatryan, Vardan and others",
    collaboration = "CMS Collaboration",
    title = "{Charged-particle nuclear modification factors in {PbPb} and {pPb} collisions at $\sqrt{s_{NN}}$=5.02 {TeV}}",
    journal = "JHEP",
    volume = "04",
    pages = "039",
    year = "2017",
    doi = "10.1007/JHEP04(2017)039",
    eprint = "1611.01664",
    archivePrefix = "arXiv",
    primaryClass = "nucl-ex"
}

@article{Bozek:2019wyr,
    author = "Bożek, Piotr and Broniowski, Wojciech and Rybczynski, Maciej and Stefanek, Grzegorz",
    title = "{{GLISSANDO} 3: {G}Lauber {I}nitial-{S}tate {S}imulation {AND} m{O}re, ver. 3}",
    eprint = "1901.04484",
    archivePrefix = "arXiv",
    primaryClass = "nucl-th",
    reportNumber = "CERN-TH-2019-011",
    doi = "10.1016/j.cpc.2019.07.014",
    journal = "Comput. Phys. Commun.",
    volume = "245",
    pages = "106850",
    year = "2019"
}

@article{ChacnAcosta2007,
  title = {Fokker-{Planck}-type equations for a simple gas and for a semirelativistic {B}rownian motion from a relativistic kinetic theory},
  volume = {76},
  journal = {Physical Review E},
  author = {Chacón-Acosta,  Guillermo and Kremer,  Gilberto M.},
  year = {2007},
  pages={02120}
}

@ARTICLE{Wilk2018,
  title     = "Sound waves in hadronic matter",
  author    = "Wilk, Grzegorz and W{\l}odarczyk, Zbigniew",
  journal   = "EPJ Web Conf.",
  volume    =  172,
  pages     = "01002",
  year      =  2018
}

@ARTICLE{Wilk2017-ta,
  title     = "Temperature oscillations and sound waves in hadronic matter",
  author    = "Wilk, G and W{\l}odarczyk, Z",
  journal   = "Physica A",
  volume    =  486,
  pages     = "579--586",
  year      =  2017
}

@article{Wilk2015,
  title = {Tsallis Distribution Decorated with Log-Periodic Oscillation},
  volume = {17},
  journal = {Entropy},
  author = {Wilk,  Grzegorz and Włodarczyk,  Zbigniew},
  year = {2015},
  pages = {384–400}
}

@article{Borsnyi2010,
  title = {The {QCD} equation of state with dynamical quarks},
  volume = {2010},
  journal = {Journal of High Energy Physics},
  author = {S. Borsányi and G. Endrődi   and Z. Fodor and A. Jakovác and S. D. Katz and S. Krieg and C. Ratti  and K. K. Szabó},
  year = {2010},
  pages={077}
}

@article{Bazavov2014,
  title = {Equation of state in 2+1 flavor {QCD}},
  volume = {90},
  journal = {Physical Review D},
  author = {Bazavov,  A. and Bhattacharya,  Tanmoy and DeTar,  C. and Ding,  H.-T. and Gottlieb,  Steven and Gupta,  Rajan and Hegde,  P. and Heller,  U. M. and Karsch,  F. and Laermann,  E. and Levkova,  L. and Mukherjee,  Swagato and Petreczky,  P. and Schmidt,  C. and Schroeder,  C. and Soltz,  R. A. and Soeldner,  W. and Sugar,  R. and Wagner,  M. and Vranas,  P.},
  year = {2014},
  pages={094503}
}

@article{Deppman2020b,
  title = {Fractal Structures of {Y}ang–{M}ills {F}ields and Non-Extensive Statistics: Applications to High Energy Physics},
  volume = {2},
  journal = {Physics},
  author = {Deppman,  Airton and Meg\'{\i}as,  Eugenio and Menezes,  Débora P.},
  year = {2020},
  pages = {455–480}
}

@article{Harris2024,
  title = {“{QGP} Signatures” revisited},
  volume = {84},
  journal = {The European Physical Journal C},
  publisher = {Springer Science and Business Media LLC},
  author = {Harris,  John W. and M\"{u}ller,  Berndt},
  year = {2024} 
}

@article{Oliveira2019,
  title     = "Complex heat capacity and entropy production of temperature modulated systems",
  author    = "de Oliveira, M{\'a}rio J.",
  journal   = "Journal of Statistical Mechanics: Theory and Experiment",
  volume    =  2019,
  pages     = "073204",
  year      = 2019
}

\end{document}